\documentclass[sort&compress,final,numberedheadings]{aipproc}
\layoutstyle{8x11double}

\begin{document}

\title{Learning the weak phase $\gamma$ from $B$ decays}

\author{Jonathan L. Rosner
\vskip -1.3in
\rightline{{\normalsize CLNS 03/1851}}
\rightline{{\normalsize hep-ph/0311170}}
\vskip 0.8in
}
{address={Laboratory of Elementary Particle Physics, Cornell University,
         Ithaca, NY 14850, U.S.A.\footnote{On leave from
Enrico Fermi Institute and Department of Physics,
University of Chicago, 5640 S. Ellis Avenue, Chicago, IL 60637.}}}

\begin{abstract}
The current status of some methods to determine the weak phase $\gamma$
of the Cabibbo-Kobayashi-Maskawa (CKM) matrix element $V^*_{ub}$ using $B$
decays is discussed, and comments are made on accuracy achievable in the next
few years.
\end{abstract}

\maketitle

\section{Introduction}

The observed CP violation in $K$ and $B$ decays can be interpreted in terms
of phases of elements of the Cabibbo-Kobayashi-Maskawa (CKM) matrix.  While
$\beta = {\rm Arg}(V^*_{td})$ is well-determined from the CP asymmetry in $B^0
\to J/\psi K_S$, current information on $\gamma = {\rm Arg}(V^*_{ub})$ is much
less precise, with $39^\circ < \gamma < 80^\circ$ at 95\% c.l.\
\cite{Hocker:2001xe}.  In order to learn $\gamma$ one must
generally separate strong and weak phases from one another in two-body $B$
decays.  We describe several areas in which progress in this work has
been accomplished, and what improvements lie ahead.  Some additional
details are noted in an earlier review \cite{Rosner:2003bq}.

In Section 2 we compare the determination of $\beta$ from
$B^0 \to J/\psi K_S$ with the more difficult determination of $\alpha = \pi
- \beta - \gamma$ from $B^0 \to \pi^+ \pi^-$.  We then discuss some uses
of information from various decay modes of $B \to K \pi$
in Sec.\ 3. One obtains useful constraints on $\gamma$ with
some assumptions about SU(3) flavor symmetry from the decays $B \to VP$
(Sec.\ 4) and $B \to PP$ (Sec.\ 5), where $V$ and $P$
denote light vector and pseudoscalar mesons.  The decays $B \to D_{CP} K$,
where $D_{CP}$ denotes a CP eigenstate of a neutral charmed meson, also
provide useful constraints (Sec.\ 6).  We summarize in Sec.\ 7.

\section{$\beta$ from $B^0 \to J/\psi K_S$ vs.\ $\alpha$ from $B^0 \to
\pi^+ \pi^-$
% \label{sec:ba}
}

The unitarity of the CKM matrix is conveniently expressed in terms of
the triangle of Fig.\ 1.  Here, for example, $1 - \bar \rho - i \bar  \eta
= - V^*_{tb} V_{td} / V^*_{cb} V_{cd}$.   (See Ref.\ \cite{Rosner:2003bq} for
other definitions.)

% This is Figure 1
\begin{figure}[h]
\includegraphics[height=0.158\textheight]{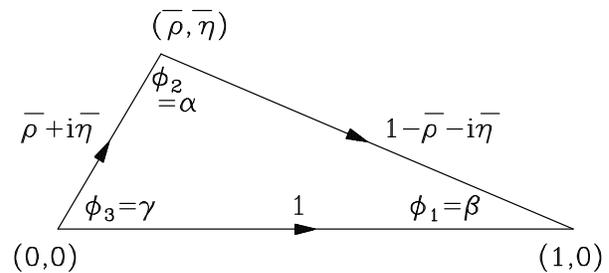}
\caption{The unitarity triangle.  Two conventions for its angles are shown.}
\end{figure}

\subsection{$B^0 \to J/\psi K_S$}

The CP asymmetry in the decay $B^0 \to J/\psi K_S$ is simple to analyze
because there is only one main subprocess $\bar b \to \bar c c \bar s$.
The direct decay (with zero weak phase) interferes with $B^0 \to \bar B^0$
mixing (with weak phase $e^{-2 i \beta}$).  The most recent BaBar and Belle
measurements, when averaged, provide $\sin2 \beta = 0.736 \pm 0.049$
\cite{Browder:2003lp} without much ambiguity.

\subsection{$B^0 \to \pi^+ \pi^-$}

Here there are two types of amplitude, ``$T$'' (tree) and ``$P$'' (penguin),
contributing to the decay.  (For a discussion of amplitudes within flavor
SU(3) see Refs.\ \cite{Gronau:1994rj} and \cite{Gronau:1995hn}.)
Different weak and strong phases can complicate the analysis.  If one had
only a tree amplitude, the direct amplitude $A(B^0 \to \pi^+ \pi^-) \sim
e^{i \gamma}$ would interfere with $A(B^0 \to \bar B^0 \to \pi^+ \pi^-)
\sim e^{- 2 i \beta} e^{-i \gamma}$ to provide a measure of the
relative phase $2(\beta + \gamma)=2 \pi - 2 \alpha$.  So, in the absence of
the penguin contribution, one would measure $\alpha$.  One seeks an estimate
of $|P/T|$ or observables not requiring this ratio.

The neutral $B$ mass eigenstates may be written
\begin{equation}
B_L^0 = p |B^0 \rangle + q | \bar B^0 \rangle~;~~
B_H^0 = p |B^0 \rangle - q | \bar B^0 \rangle~,
\end{equation}
where
\begin{eqnarray}
q/p & = & e^{- 2 i \beta}~,~~\lambda \equiv (q/p)(\bar A / A)~, \\
A & \equiv & A(B \to f)~,~~\bar A \equiv A(\bar B \to \bar f)~.
\end{eqnarray}
Observables in the time-dependence of
$\left\{ \begin{array}{c}B^0 \\ \bar B^0 \end{array} \right\}_{t=0} \to
\pi^+ \pi^-$ (or any other final state) are:
\begin{equation}
\Gamma(t) \sim e^{- \Gamma |t|}
[1 \mp S \sin \Delta m t \mp A \cos \Delta m t]~,~~t \equiv t_{\rm decay}
- t_{\rm tag}~,
\end{equation}
with $S = 2 {\rm Im} \lambda/(1 + |\lambda|^2)$ and
\begin{equation}
A = \frac{|\lambda|^2 - 1}{|\lambda|^2 + 1} = A_{CP}
= \frac{\Gamma(\bar B \to \bar f) - \Gamma(B \to f)}
{\Gamma(\bar B \to \bar f) + \Gamma(B \to f)}~.
\end{equation}
The experimental data \cite{Jawahery:2003lp,Chiang:2003pp}
on these asymmetries in $B \to \pi^+ \pi^-$ are shown in Table 1.

% This is Table 1
\begin{table}
\begin{tabular}{cccc} \hline
\tablehead{1}{c}{b}{Observable}
& \tablehead{1}{c}{b}{BaBar}
& \tablehead{1}{c}{b}{Belle}
& \tablehead{1}{c}{b}{Average} \\ \hline
$S_{\pi \pi}$ & $-0.40 \pm 0.22 \pm 0.33$ &
$-1.23 \pm 0.41^{+0.08}_{-0.07}$ & $-0.58 \pm 0.20$ \\
$A_{\pi \pi}$ & $0.19 \pm 0.19 \pm 0.05$ &
$0.77 \pm 0.27 \pm 0.08$ & $0.38 \pm 0.16$ \\ \hline
\end{tabular}
\caption{Time-dependent asymmetries in $B^0 \to \pi^+ \pi^-$.}
\label{tab:tda}
\end{table}

With no penguin contributions, $S_{\pi \pi} = \sin 2 \alpha < 0$ would
favor $\alpha > 90^\circ$.  With a penguin-to-tree ratio $|P/T| \simeq 0.3$
estimated from $B \to K \pi$
using flavor SU(3) symmetry, one finds instead the parametric dependence
of the time-dependent asymmetries on $\alpha$ and a relative strong phase
$\delta$ \cite{Gronau:2002qj,Gronau:2002gj} as shown in Fig.\ 2
\cite{Gronau:2003cq}.

% This is Figure 2
\begin{figure}[h]
\includegraphics[height=.18\textheight]{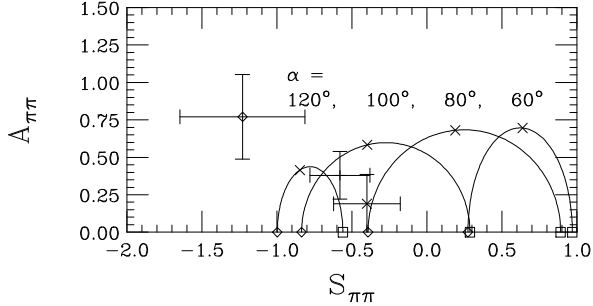}
\caption{Curves describing behavior of $S_{\pi \pi}$ and $A_{\pi \pi}$ as
the relative strong phase $\delta$ between penguin and tree amplitudes is
varied from $0^\circ$ (diamonds) through $90^\circ$ (crosses) to
$180^\circ$ (squares).  The curves are labeled by values of $\alpha$.
Plotted points:  Babar (cross), Belle (diamond), average (no symbol).} 
\end{figure}

Unless $\delta$ is near $\pi/2$, curves for different
values of $\alpha$ intersect at the same values of $S_{\pi \pi}$ and
$A_{\pi \pi}$.  A quantity which is useful in resolving this discrete
ambiguity is $R_{\pi \pi} = \frac{\Gamma(B^0 \to \pi^+ \pi^-)}
{\Gamma(B^0 \to \pi^+ \pi^-)_{\rm tree}}$.  In Ref.\ \cite{Luo:2003hn}
it was found that $R_{\pi \pi}= 0.87^{+0.11}_{-0.28}$, which slightly favors
larger strong phases and hence larger values of $\alpha$ for given
$(S_{\pi \pi},A_{\pi \pi})$.  A related analysis has appeared recently in
Ref.\ \cite{Buchalla:2003jr}.  Information on $B_s \to K^+ K^-$ may be
combined with that on $B \to \pi^+ \pi^-$ with the help of flavor SU(3) to
separate out penguin and tree contributions \cite{Fleischer:1999pa}.
The time-dependence of $B_s(t) \to K^+ K^-$ provides a complementary
method \cite{Gronau:2002bh}.

\section{Information from $B \to K \pi$
% \label{sec:kpi}
}

A great deal of information can be obtained from $B \to K \pi$ decay rates
averaged over CP, supplemented with measurements of direct CP asymmetries.
One probes in this manner tree-penguin interference in various processes.
Denoting amplitudes with $|\Delta S|=1$ by primed quantities, several
comparisons can be made:
\begin{itemize}

\item $B^0 \to K^+ \pi^-~(T'+P'$) vs.\ $B^+ \to K^0 \pi^+~(P'$) \\
\cite{Fleischer:1998um,Gronau:1998an,Gronau:2001cj,Gronau:2003kj};

\item $B^+ \to K^+ \pi^0~(T'+P'+C')$ vs.\ $B^+ \to K^0 \pi^+~(P')$
\cite{Gronau:2003kj,Neubert:1998pt,Neubert:1998jq,Neubert:1998re};

\item $B^0 \to K^0 \pi^0$ vs.\ other modes \cite{Gronau:2003kj,Buras:1998rb,%
Buras:2000gc,Beneke:2001ev,Beneke:2002jn,Beneke:2003zv}.

\end{itemize}
The data which are used in these analyses are summarized in Table 2.
% This is Table 2
\begin{table}
\caption{Branching ratios and CP asymmetries for $B \to K \pi$ decays
\cite{Chiang:2003pp}.}
\begin{tabular}{c c c c} \hline
Decay mode & Amplitude & ${\cal B}$ (units of $10^{-6}$) & $A_{CP}$ \\ \hline
$B^+ \to K^0 \pi^+$ & $P'$ & $21.78 \pm 1.40$ & $0.016 \pm 0.057$ \\
$B^+ \to K^+ \pi^0$ & $-(P'+C'+T')/\sqrt{2}$ & $12.53 \pm 1.04$ &
 $0.00 \pm 0.12$ \\
$B^0 \to K^+ \pi^-$ & $-(T'+P')$ & $18.16 \pm 0.79$ & $-0.095 \pm 0.029$ \\
$B^0 \to K^0 \pi^0$ & $(P'-C')/\sqrt{2}$ & $11.68 \pm 1.42$ &
 $0.03 \pm 0.37$ \\
\hline
\end{tabular}
\end{table}

In all these comparisons it is helpful to use flavor SU(3) (often only U-spin,
i.e., $s \leftrightarrow d$).  We give the example of $B^0 \to K^+ \pi^-$
in detail.  The tree amplitude for this process is $T' \sim V_{us}V^*_{ub}$,
with weak phase $\gamma$, while the penguin amplitude is $P' \sim V_{ts}
V^*_{tb}$ with weak phase $\pi$.  We denote the penguin-tree relative strong
phase by $\delta$ and define $r \equiv |T'/P'|$.  Then we may write
\begin{eqnarray}
A(B^0 \to K^+ \pi^-) & = & |P'|[1 - r e^{i (\gamma + \delta)}]~, \\
A(\bar B^0 \to K^- \pi^+) & = & |P'|[1 - r e^{i (-\gamma + \delta)}]~, \\
A(B^+ \to K^0 \pi^+) & = & A(B^- \to \bar K^0 \pi^-) = - |P'|~,
\end{eqnarray}
where the last two amplitudes are expected to be equal in the approximation
that small annihilation amplitudes are neglected.  A test for this assumption
is the absence of a CP asymmetry in $B^+ \to K^0 \pi^+$ (or in $B^+ \to \bar
K^0 K^+$, where it would be bigger \cite{Falk:1998wc}).

One now forms the ratio
\begin{eqnarray}
R & \equiv & \frac{\Gamma(B^0 \to K^+ \pi^-) + \Gamma(\bar B^0 \to K^- \pi^+)}
{2\Gamma(B^+ \to K^0 \pi^+)} \nonumber \\
 & = & 1 - 2 r \cos \gamma \cos \delta + r^2~.
\end{eqnarray}
Fleischer and Mannel \cite{Fleischer:1998um} pointed out that $R \ge \sin^2
\gamma$ for any $r, \delta$ so if $1 > R$ one can get a useful bound.  However,
if one uses
\begin{equation}
R A_{CP} = - 2 r \sin \gamma \sin \delta
\end{equation}
as well and eliminates $\delta$ one can get a more powerful constraint,
illustrated in Fig.\ 3.

% This is Figure 3
\begin{figure}
\includegraphics[height=.33\textheight]{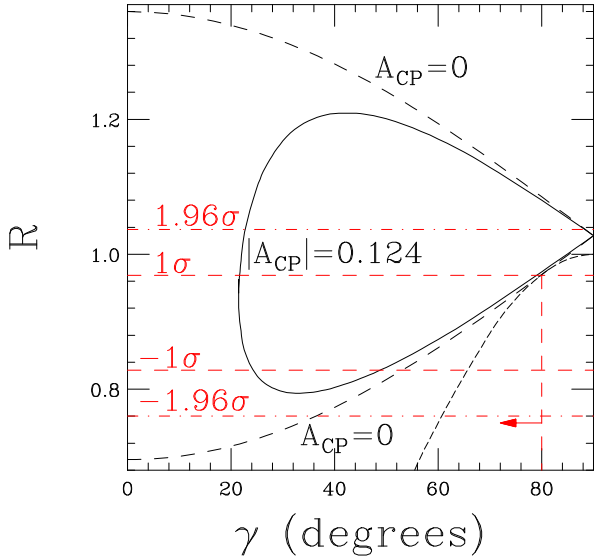}
\caption{Behavior of $R$ for $r = 0.166$ and $A_{CP} = 0$ (dashed curves) or
$|A_{CP}| = 0.124$ (solid curve) as a function of the weak phase $\gamma$.
Horizontal dashed lines denote $\pm 1 \sigma$ experimental limits on $R$,
while dot-dashed lines denote $95\%$ c.l. ($\pm 1.96 \sigma$) limits.  The
short-dashed curve denotes the Fleischer-Mannel bound $\sin^2 \gamma \le R$.}
\end{figure}

At the $1 \sigma$ level, $R < 1$, leading to an upper bound $\gamma <
80^\circ$ which happens to coincide with that of Ref.\ \cite{Fleischer:1998um}.
We have used $R = 0.898 \pm 0.071$ and $A_{CP} = -0.095 \pm 0.029$ based on
recent averages \cite{Chiang:2003pp} of CLEO, BaBar, and Belle data, and $r =
|T'/P'| = 0.142^{+0.024}_{-0.012}$.  The most conservative bound arises for
the smallest $|A_{CP}|$ and largest $r$.  The allowed region lies between the
curves $A_{CP} = 0$ and $|A_{CP}| = 0.124~(1 \sigma)$.  In order to estimate
the tree amplitude and $r = |T'/P'|$ we have used factorization in $B^+ \to
\pi^- \ell^+ \nu_\ell$ at low $q^2$ \cite{Luo:2003hn} and $\left|\frac{T'}
{T} \right| = \frac{f_K}{f_\pi}\left| \frac{V_{us}}{V_{ud}} \right| \simeq
(1.22)(0.23) = 0.28$.  One could use processes in which $T$ dominates, such as
$B^0 \to \pi^+ \pi^-$ or $B^+ \to \pi^+ \pi^0$, but these are contaminated by
contributions from $P$ and $C$, respectively.

In such an approach one always must question the validity of SU(3) flavor
symmetry.  SU(3) breaking is taken into account in the ratio of tree
amplitudes, but no breaking is taken in other amplitudes, since we do not
assume factorization for $C$ or $P$ and therefore cannot account for the
breaking merely via ratios of decay constants.  We have assumed the same
relative tree-penguin strong phases for $|\Delta S| = 1$ and $\Delta S = 0$
amplitudes.  Tests of these assumptions will be available once one observes
penguin-dominated $B \to K \bar K$ decays and charmless $B_s$ decays; there
are also numerous relations implied between CP-violating rate differences
\cite{Deshpande:1995ii,Gronau:2000zy}.

The process $B^+ \to K^+ \pi^0$ also provides constraints on $\gamma$.  The
deviation of the ratio
\begin{equation}
R_c \equiv \frac{\Gamma(B^+ \to K^+ \pi^0) + \Gamma(B^- \to K^- \pi^0)}
{\Gamma(B^+ \to K^0 \pi^+)} =1.15 \pm 0.12
\end{equation}
from 1, when combined with $A_{CP} = 0.00 \pm 0.12$, $r_c = |(T'+C')/P'| =
0.195\pm 0.016$ and an estimate of the electroweak penguin (EWP) $\delta_{EW}
\equiv |P'_{EW}|/ |T'+C'| = 0.65 \pm 0.15$, leads to a $1 \sigma$ lower bound
$\gamma > 40^\circ$.  Details may be found in Refs.\ \cite{Rosner:2003bq,%
Gronau:2003kj,Neubert:1998pt,Neubert:1998jq,Neubert:1998re}.
The most conservative bound arises for smallest $A_{CP}$, largest $r_c$,
and largest $|P'_{EW}|$, and is shown in Fig.\ 4.

% This is Figure 4
\begin{figure}
\includegraphics[height=.328\textheight]{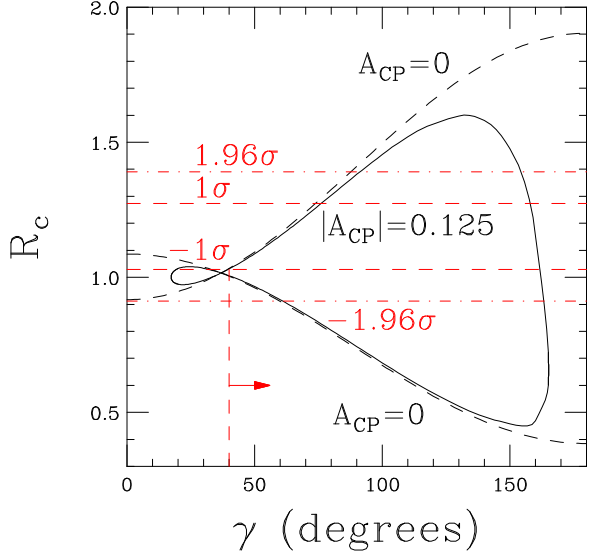}
\caption{Behavior of $R_c$ for $r_c = 0.21$ ($1 \sigma$ upper limit) and
$A_{CP}(K^+ \pi^0) = 0$ (dashed curves) or $|A_{CP}(K^+ \pi^0)|
= 0.125$ (solid curve) as a function of the weak phase $\gamma$.
Horizontal dashed lines denote $\pm 1 \sigma$ experimental limits on $R_c$,
while dotdashed lines denote 95\% c.l. ($ \pm 1.96 \sigma$) limits.  We have
taken $\delta_{EW} = 0.80$ (its $1 \sigma$ upper limit), which
leads to the most conservative bound on $\gamma$.}
\end{figure}

Another ratio
\begin{eqnarray}
R_n & \equiv & \frac{\bar \Gamma(B^0
\to K^+ \pi^-)}{2 \bar \Gamma(B^0 \to K^0 \pi^0)} \nonumber \\
& = & \left| \frac{p'+t'} {p'-c'} \right|^2 = 0.78 \pm 0.10
\end{eqnarray}
involves the decay $B^0 \to K^0 \pi^0$.  Here the bar denotes CP-averaged
decay widths, while small letters denote amplitudes which include EWP
contributions.  This ratio should be same to leading order in $|t'/p'|$ and
$|c'/p'|$ as
\begin{equation}
R_c = \left| \frac{p'+t'+c'}{p'} \right|^2~~,
\end{equation}
but the two ratios differ by $2.4 \sigma$.  Possibilities (see, e.g., Refs.\
\cite{Gronau:2003kj,Grossman:2003lp})
include (1) new physics, e.g., in the EWP amplitude, and (2) an underestimate
of the $\pi^0$ detection efficiency in all experiments, leading to an
overestimate of any branching ratio involving a $\pi^0$.  The latter
possibility can be taken into account by considering the ratio
$(R_n R_c)^{1/2} = 0.96 \pm 0.08$, in which the $\pi^0$ efficiency cancels.
As shown in Fig.\ 5, this ratio leads only to the conservative bound
$\gamma \le 88^\circ$.

% This is Figure 5
\begin{figure}
\includegraphics[height=.32\textheight]{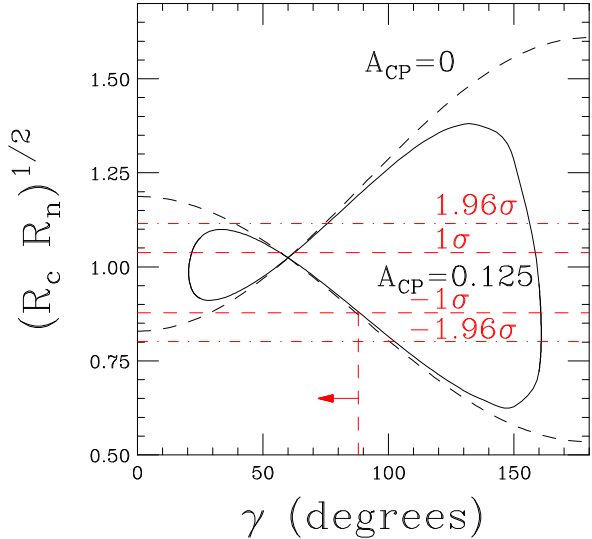}
\caption{Behavior of $(R_c R_n)^{1/2}$ for $r_c = 0.18$ ($1 \sigma$ lower
limit) and $A_{CP}(K^+ \pi^0) = 0$ (dashed curves) or
$|A_{CP}(K^+ \pi^0)| = 0.125$ (solid curve) as a function of the
weak phase $\gamma$. Horizontal dashed lines denote $\pm 1 \sigma$ experimental
limits on $(R_c R_n)^{1/2}$, while dotdashed lines denote 95\% c.l. ($ \pm 1.96
\sigma$) limits.  Upper branches of curves correspond to $\cos \delta_c(\cos
\gamma - \delta_{EW}) < 0$, while lower branches
correspond to $\cos \delta_c(\cos \gamma - \delta_{EW}) > 0$.  Here we have
taken $\delta_{EW} = 0.50$ (its $1 \sigma$ lower limit), which
leads to the most conservative bound on $\gamma$.}
\end{figure}

\section{Information from $B \to VP$
% \label{sec:vp}
}

Although the decays $B \to VP$ are characterized by more amplitudes than $B
\to PP$ (since the final particles do not belong to the same flavor-symmetry
multiplet), data have become so abundant that useful global fits can be
performed \cite{Aleksan:2003qi,Chiang:2003pm}. We label amplitudes by the meson
(pseudoscalar $P$ or vector $V$) containing the spectator quark.  Some
features of the fit of Ref.\ \cite{Chiang:2003pm} are:

\begin{itemize}

\item $|t_P/t| \simeq f_\rho/f_\pi$, where $t$ is the tree amplitude in
$B \to PP$ decays, as would be expected for a weak current
producing a charged meson.

\item Penguin amplitudes satisfy $p'_V \simeq - p'_P$, as proposed long ago
by Lipkin \cite{Lipkin:1991us,Lipkin:1997ad,Lipkin:1998ew}.

\item Small CP asymmetries in many processes imply small strong phases.

\item The time-dependent asymmetries in $B \to \rho \pi$ are crucial in
resolving discrete ambiguities, as has also been found in the QCD factorization
approach of Refs.\ \cite{Beneke:2001ev,Beneke:2002jn,Beneke:2003zv}.

\end{itemize}

Three local $\chi^2$ minima are found:  $\gamma = (26 \pm 5)^\circ$, $(63 \pm 6)
^\circ$ (a range compatible with fits \cite{Hocker:2001xe} to CKM parameters),
and $(162^{+5}_{-6})^\circ$ (incompatible with $\beta \simeq 24^\circ$ since
$\alpha + \beta + \gamma = \pi$).  At 95\% c.l. the solution compatible with
CKM fits implies $51^\circ \le \gamma \le 73^\circ$ and small strong
phases in accord with the expectations of QCD factorization
\cite{Beneke:2003zv}.  Some
predictions for as-yet-unseen decay modes are shown in Table 3.

% This is Table 3
\begin{table}
\caption{Predictions of the favored fit of Ref.\ \cite{Chiang:2003pm} for some
as-yet-unseen $B \to V P$ decays.}
\begin{tabular}{c c c c} \hline
As yet unseen & Predicted ${\cal B}$ & Present limit   & Comments \\
decay mode   & \multicolumn{2}{c}{Units of $10^{-6}$}  &      \\ \hline
$B^+ \to \bar K^{*0} K^+$ & $0.50 \pm 0.05$  & $< 5.3$ & Pure $p_P$ \\
$B^+ \to K^{*+} \pi^0$ & $15.0^{+3.3}_{-2.8}$ & $< 31$ & EWP enhancement  \\
$B^+ \to \rho^+ K^0$   & $12.6 \pm 1.6$       & $< 48$ & Pure $p'_V$ \\
$B^0 \to \rho^0 K^0$   & $7.2^{+2.1}_{-1.9}$  & $< 12.4$ & EWP enhancement \\
\hline
\end{tabular}
\end{table}

In this fit the free parameters were:

\begin{itemize}

\item $p'_{P,V}$ (penguin amplitudes); their relative phase $\phi$;

\item  $t_{P,V}$ (tree amplitudes); their strong
phases $\delta_{P,V}$ with respect to $p_{P,V}$;

\item Color-suppressed $c_{P,V}$ taken real with respect to $t_{P,V}$;

\item Electroweak penguins $P'_{EW(P,V)}$ taken real with respect
to $p'_{P,V}$;

\item The weak phase $\gamma$.

\end{itemize}

One thus has 12 parameters (11 if $p'_V/p'_P$ is assumed to be real, and
10 if we assume $p'_V/p'_P = -1$) to fit 34 data points.  The resulting
dependence of $\chi^2_{\rm min}$ on $\gamma$ is shown in Fig.\ 6.

% This is Figure 6
\begin{figure}
\includegraphics[height=.28\textheight]{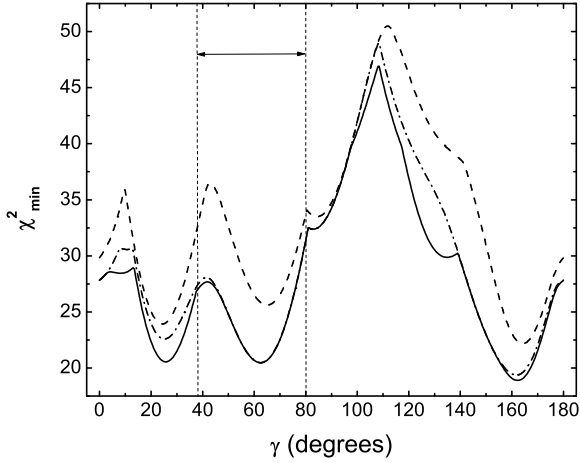}
\caption{$(\chi^2)_{\rm min}$, obtained by minimizing over all remaining fit
parameters, as a function of the weak phase $\gamma$.  Dashed curve: $p'_V/p'_P
= -1$ (24 d.o.f.); dash-dotted curve: $p'_V/p'_P$ real (23 d.o.f.); solid
curve: $p'_V/p'_P$ complex (22 d.o.f.).  Vertical dashed lines show the limits
$39^\circ \le \gamma \le 80^\circ$ from fits \cite{Hocker:2001xe} to CKM
parameters.}
\end{figure}

The relative phases of amplitudes are specified by CP-averaged decay rates
as well as by CP asymmetries.  They are illustrated in Fig.\ 7.

% This is Figure 7
\begin{figure}
\includegraphics[height=.36\textheight]{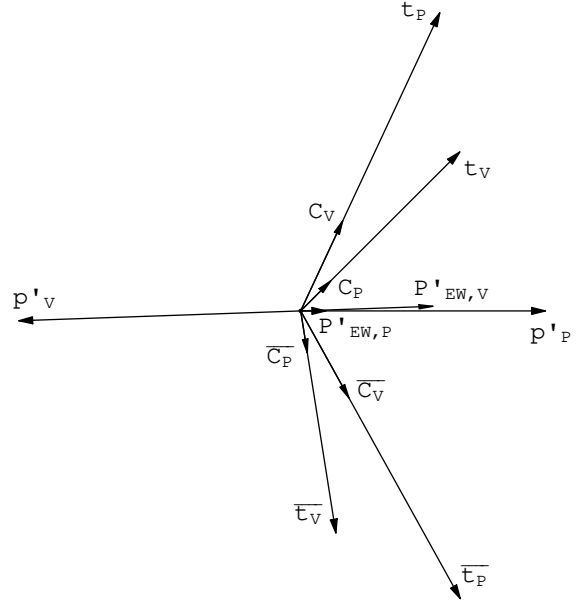}
\caption{Magnitudes and phases of dominant invariant amplitudes in solution
with $\gamma \simeq 63^\circ$ and complex $p'_V/p'_P$ \cite{Chiang:2003pm}.}
\end{figure}

The $p'_P$ amplitude in this diagram is taken to be real and positive.  The
weak phases of $t_{V,P}$ and $\bar t_{V,P}$ are included.  There is a small
relative phase between $t_V$ and $t_P$, as expected in QCD factorization
\cite{Beneke:2003zv}.  The relative phases of $p'_V$ and $p'_P$ are such that
they contribute constructively to $B \to K^* \eta$, as anticipated by
Lipkin \cite{Lipkin:1991us,Lipkin:1997ad,Lipkin:1998ew}.  From these phases
one expects constructive tree-penguin interference in the CP-averaged
rate for $B^0 \to K^{*+} \pi^-$ and destructive interference in $B^0 \to
K^+ \rho^-$.

\section{$B \to PP$ decays with $\eta$ and $\eta'$
% \label{sec:pp}
}
A global fit to $B \to PP$ decays under the same assumptions as the fit to $B
\to VP$ decays is still in progress \cite{Chiang:2003pp}.  However, $B \to PP$
decays with $\eta$ and $\eta'$ have been analyzed using flavor symmetry in
Ref.\ \cite{Chiang:2003rb}.  It is found that the large CP asymmetry in $B^+
\to \pi^+ \eta$ reported by the BaBar Collaboration \cite{Aubert:2003ez}
implies a comparable $A_{CP}$ in $B^+ \to \pi^+ \eta'$.  We have
\begin{equation}
A_{CP}(\pi^+ \eta) = - \frac{0.91 \sin \alpha
\sin \delta}{1 - 0.91 \cos \alpha \cos \delta} = -0.51 \pm 0.19~~,
\end{equation}
\begin{eqnarray}
\bar {\cal B}(\pi^+ \eta) & = & 4.95 \times 10^{-6}(1 - 0.91 \cos
\alpha \cos \delta) \nonumber \\ & = & (4.12 \pm 0.85) \times 10^{-6}~~,
\end{eqnarray}
leading to the predictions
\begin{equation}
A_{CP}(\pi^+ \eta') = - \frac{\sin \alpha
\sin \delta}{1 - \cos \alpha \cos \delta} \simeq -0.57~~,
\end{equation}
\begin{eqnarray}
\bar {\cal B}(\pi^+ \eta') & = & 3.35 \times 10^{-6}(1 - \cos \alpha \cos
\delta) \nonumber \\
& \simeq & 2.7 \times 10^{-6}~~.
\end{eqnarray}
Tree and penguin amplitudes are of comparable magnitude
in both these processes, leading to the possibility of large CP asymmetries
which appears to be realized in the data.  The scatter of predictions for
$B^+ \to \pi^+ \eta'$ is shown in Fig.\ 8.  The central values are based on
$(\alpha,\delta) \simeq (78^\circ,28^\circ)$ with the discrete ambiguities
$(\alpha \leftrightarrow \delta)$ or $\alpha \to \pi - \alpha$,
$\delta \to \pi - \delta$.

% This is Figure 8
\begin{figure}[t]
\includegraphics[height=.36\textheight]{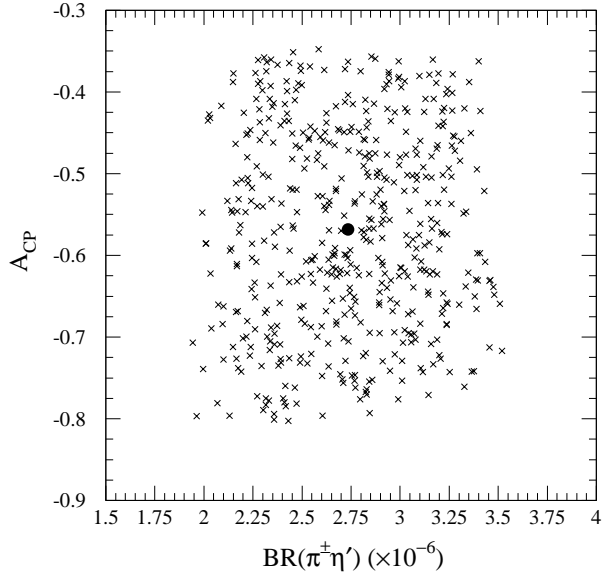}
\caption{Predicted values of the averaged branching ratio and direct $CP$
  asymmetry for the decays $B^\pm \to \pi^\pm \eta'$.}
\end{figure}

\section{Constraints from $B \to D_{CP} K$
% \label{sec:DK}
}

We present the discussion of M. Gronau \cite{Gronau:2003cq}.
One wishes to compare $B^\pm \to D_{CP}K^\pm$ with the
CKM-favored process $B^- \to D^0 K^-$, thereby probing the quark subprocesses
\begin{equation}
\frac{A(\bar b \to \bar u c \bar s)}{A(\bar b \to \bar c u \bar s)}
 =  r e^{i(\gamma + \delta)}~,~~\frac{A(b \to u \bar c s)}{A(b \to c \bar u s)}
 = r e^{i(-\gamma + \delta)}~~.
\end{equation}
One then considers the ratios
\begin{eqnarray}
R_\pm & \equiv & \frac{\Gamma(D^0_{CP=\pm}K^-) + \Gamma(D^0_{CP=\pm}K^+)}
{\Gamma(D^0 K^-)} \nonumber \\
& = & 1 + r^2 \pm 2 r \cos \gamma \cos \delta
\end{eqnarray}
and the CP asymmetries
\begin{eqnarray}
A_\pm & \equiv & \frac{\Gamma(D^0_{CP=\pm}K^-) - \Gamma(D^0_{CP=\pm}K^+)}
{\Gamma(D^0_{CP=\pm}K^-) + \Gamma(D^0_{CP=\pm}K^+)} \nonumber \\
& = & \pm 2 r \sin \gamma \sin \delta / R_{\pm}~~.
\end{eqnarray}
The relevant data are summarized in Table 4.  The ratio $\Gamma(B^- \to D^0
K^-)/\Gamma(B^- \to D^0 \pi^-)$ was evaluated \cite{Gronau:2003cq} by taking
the average of CLEO \cite{Bornheim:2003bv}, Belle \cite{Swain:2003yu}, and
BaBar \cite{Aubert:2003uy} values.

% This is Table 4.
\begin{table}
\caption{Ratios $R_\pm$ and CP asymmetries $A_\pm$ for $B \to D_{CP}K$ decays.}
\begin{tabular}{c c c c c} \hline
        & $R_+$ & $A_+$ & $R_-$ & $A_-$ \\ \hline
Belle \cite{Swain:2003yu}  & $1.12 \pm 0.24$ & $0.06 \pm 0.19$ &
 $1.30 \pm 0.25$ & $-0.19 \pm 0.18$ \\
BaBar \cite{Aubert:2003uy} & $1.06 \pm 0.21$ & $0.07 \pm 0.18$ & & \\
Average                    & $1.09 \pm 0.16$ & $0.065 \pm 0.132$ &
 $1.30 \pm 0.25$ & $-0.19 \pm 0.18$ \\
\hline
\end{tabular}
\end{table}
We take $(R_+ + R_-)/2 = 1+r^2$ so
$r \ge 0.22$ at $1 \sigma$.  The average $|A_\pm|$ is $0.11 \pm 0.11
\le 0.22$ at $1 \sigma$; the most conservative bound is obtained for smallest
$r$ and largest $|A_\pm|$ and at $1 \sigma$ is $\gamma \ge 72^\circ$, as shown
in Fig.\ 9.  Note that one $R$ must be below $1 + r^2$ while the other must be
above it.

% This is Figure 9
\begin{figure}
\includegraphics[height=.33\textheight]{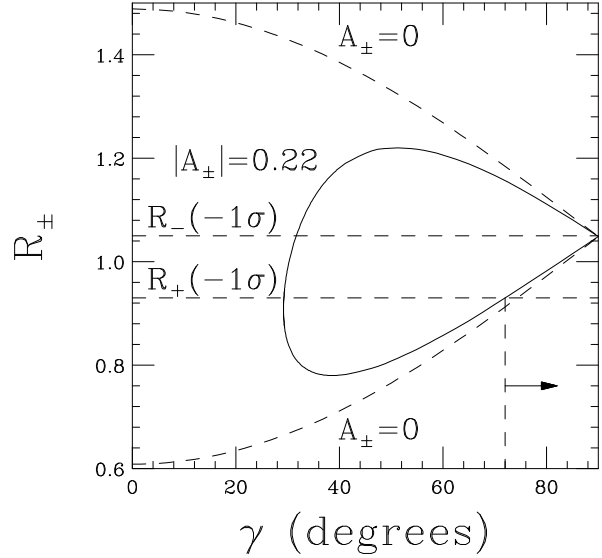}
\caption{Behavior of $R_{\pm}$ for
$A_{\pm} = 0$ (dashed curves) or $|A_{\pm}| = 0.22$ (solid curve) as a
function of the weak phase $\gamma$.  Horizontal dashed lines denote $-1
\sigma$ experimental limits on $R_\pm$.  We have taken parameters
(including $r=0.22$) which lead to the most conservative bound on $\gamma$.}
\end{figure}

\section{Summary
% \label{sec:sum}
}
A number of promising bounds on $\gamma$ stemming from various $B$ decays have
been mentioned.  So far all are statistics-limited.  At $1 \sigma$ we have
found

\begin{itemize}

\item $R$ ($K^+ \pi^-$ vs.\ $K^0 \pi^+$) gives $\gamma \le 80^\circ$;
 
\item $R_c$ ($K^+ \pi^0$ vs.\ $K^0 \pi^+$) gives $\gamma \ge 40^\circ$;

\item $R_n$ ($K^+ \pi^-$ vs.\ $K^0 \pi^0$) should equal $R_c$;
$(R_c R_n)^{1/2}$ gives $\gamma \le 88^\circ$;

\item $B \to D_{CP} K$ decays give $\gamma \ge 72^\circ$.

\end{itemize}

A flavor-SU(3) analysis of $B \to VP$ decays favors $\gamma = (63 \pm 6)
^\circ$, or $51^\circ \le \gamma \le 73^\circ$ at 95\% c.l.\  Several
as-yet-unseen decay modes are predicted, such as $B^+ \to \rho^+ K^0$ and $B^+
\to K^{*+} \pi^0$.  SU(3) relations among rate differences remain to be
tested.  We predict $2.0 \le \bar {\cal B}(\pi^+ \eta')/10^{-6} \le 3.5$,
$-0.34 \ge A_{CP}(\pi^+ \eta') \ge -0.80$.  A global $B \to PP$ analysis,
still in progress, is complicated by possible $B \to K \pi$ inconsistencies or
new physics in (e.g.) $B^0 \to K^0 \pi^0$.

The future of most such $\gamma$ determinations remains for now in
experimentalists' hands, as one can see from Figs.\ 3--5 and 9.  Uncertainties
in SU(3) breaking are probably already the limiting factor on the error in
$\gamma$ from Fig.\ 6, and better estimates will require flavor SU(3) tests at
levels of ${\cal B} \simeq 1/2 \times 10^{-6}$.  We have noted
(see, e.g., \cite{Gronau:1998an}) that measurements of rate ratios
in $B \to K \pi$ can ultimately pinpoint $\gamma$ to within about $10^\circ$.
The required accuracies in $R$, $R_c$, and $R_n$ to achieve this goal can be
estimated from Figs.\ 3--5.  For example, knowing $(R_c R_n)^{1/2}$ to within
0.05 would pin down $\gamma$ to within $10^\circ$ if this ratio lies in the
most sensitive range of Fig.\ 5.

A complementary approach to the flavor-SU(3) method is the QCD
factorization formalism of Refs.\ \cite{Beneke:2001ev,Beneke:2002jn,%
Beneke:2003zv}.  It predicts small strong phases (as
found in our analysis) and deals directly with flavor-SU(3) breaking; however,
it involves some unknown form factors and meson wave functions and appears
to underestimate the magnitude of $B \to VP$ penguin amplitudes.

\begin{theacknowledgments}
I would like to thank Cheng-Wei Chiang, Michael Gronau, Zumin Luo, Matthias
Neubert, and Denis Suprun for enjoyable collaborations on the material
presented here.  This work was supported in part by the United
States Department of Energy through Grant No.\ DE FG02 90ER40560.
\end{theacknowledgments}

\bibliographystyle{aipproc}

\bibliography{gamma}

\IfFileExists{\jobname.bbl}{}
 {\typeout{}
  \typeout{******************************************}
  \typeout{** Please run "bibtex \jobname" to optain}
  \typeout{** the bibliography and then re-run LaTeX}
  \typeout{** twice to fix the references!}
  \typeout{******************************************}
  \typeout{}
 }

\end{document}